\documentclass[a4paper,11pt]{article}
\usepackage{pos}
\usepackage{amsmath,amsfonts,dsfont}
\renewcommand{\logo}{\relax}

\title{The pion-nucleon sigma term with $N_f=2+1$ $\mathcal{O}(a)$-improved Wilson fermions}

\author[a]{A.~Agadjanov}
\author*[b,c]{D.~Djukanovic}
\author[a]{G.~von~Hippel}
\author[a,b,c]{H.B.~Meyer}
\author[a]{K.~Ottnad}
\author[a,b,c]{H.~Wittig}

\affiliation[a]{PRISMA$^+$ Cluster of Excellence \& Institut f\"ur Kernphysik,
Johannes Gutenberg-Universit\"at Mainz, D-55099 Mainz, Germany}
\affiliation[b]{Helmholtz Institute Mainz, Staudingerweg 18, D-55128 Mainz, Germany}
\affiliation[c]{GSI Helmholtzzentrum f\"ur Schwerionenforschung, D-64291
Darmstadt}

\emailAdd{d.djukanovic@him.uni-mainz.de}

\abstract{
We present an analysis of the pion-nucleon sigma term on the CLS ensembles with
$N_f=2+1$ flavors of $\mathcal{O}(a)$-improved Wilson fermions. We perform a chiral
interpolation based on ensembles with pion masses ranging from 130 MeV to
roughly 350 MeV. The analysis covers four lattice spacings between
$a\approx [0.05 \, \rm{fm}\ldots 0.09\, \rm{fm}]$, allowing for an estimate of systematics associated with lattice
artefacts.}

\FullConference{%
The 39th International Symposium on Lattice Field Theory,\\
8th-13th August, 2022,\\
Rheinische Friedrich-Wilhelms-Universität Bonn, Bonn, Germany
}

\begin{document}
\maketitle

\section{Introduction}
The pion-nucleon sigma term is defined as the matrix element of the scalar
current
\begin{align}
	\sigma_{\pi N}&= {m_{l}} \langle N | \bar u\,  u +\bar d \, d | N \rangle
	\label{eq:def_sigma}
\end{align}
where $u$ and $d$ denote the up- and down-quark, $m_l$ is the light quark mass, inserted between nucleon states
$|N\rangle$. 
Alternatively the sigma term is accessible via the Feynman-Hellmann theorem
\cite{feynman:1939zza,hellmann:1933}, that relates
$\sigma_{\pi N}$ to derivatives of the nucleon mass with respect to the light
quark masses $m_\ell=(m_u+m_d)/2$, i.e.
\begin{align}
	\sigma_{\pi N}&=  \frac{\partial}{\partial m_\ell} m_N  .
	\label{eq:fh}
\end{align}
The sigma term is of particular phenomenological interest, as it
proves vital in constraining beyond
standard model physics. More specifically, a promising class of dark matter
candidates, the so-called weakly interacting massive particles, may leave an
imprint in scattering with ordinary nuclear matter via scalar interactions. The
contribution of the spin-independent cross
section to the rate of the WIMP-nucleus scattering is enhanced by the number of
nuclei within the nucleus \cite{Ellis:2008hf}. Given the pion-nucleon sigma term one thus may infer bounds on
the masses of WIMPs via direct detection experiments measuring the recoil energy
of heavy nuclei scattered on WIMPs, e.g. Xenon \cite{XENON:2018voc} or CDMS
\cite{CDMS-II:2010iir}.

Despite the lack of a scalar probe for experiments, there still is a
connection to an experimentally accessible quantity, i.e. pion nucleon
scattering lengths.
This connection is established via the Cheng-Dashen low energy theorem
\cite{Cheng:1970mx}, that relates the scattering amplitudes from pion-nucleon
scattering at subthreshold kinematics, to the scalar form factor. In the
framework of Roy-Steiner equations the following value for the pion-nucleon
sigma term has been obtained
\cite{Hoferichter:2015dsa} 
\begin{align}
	\sigma_{\pi N}&=  (59.1 \pm 3.5) \, \rm{MeV}.
	\label{eq:sigma_rs}
\end{align}
However, comparing this phenomenological value to recent
determinations from the lattice
\cite{Alexandrou:2019brg,Gupta:2021ahb,Yang:2015uis,Yamanaka:2018uud,Borsanyi:2020bpd,Bali:2016lvx}, 
a slight tension emerges, where
lattice extractions tend to smaller values. Most recently, in Ref. \cite{Gupta:2021ahb} it was argued that this
discrepancy might be alleviated with a careful treatment of excited states
in the direct determination.

In this proceedings contribution, we report on progress in extracting the pion-nucleon sigma
term on CLS ensembles with $N_f=2+1$ $\mathcal{O}(a)$ improved Wilson fermions
\cite{Bruno:2014jqa}. 
We focus on the direct extraction, calculating the matrix element of
Eq.~(\ref{eq:def_sigma}).

\section{Lattice Methodology}
The Wick contractions for the scalar matrix element of
Eq.~(\ref{eq:def_sigma}) generate connected and quark-disconnected contributions. For
the connected contributions we apply the standard methods, as described in \cite{Djukanovic:2021cgp}, where sequential
propagators are calculated in a fixed-sink setup. We calculate the two-point
function  and the 
three-point function of the scalar current
\begin{align}
	C_2(t,\vec p)&=\Gamma_{\alpha \beta}\sum\limits_{\vec x} e^{-i\vec p\vec x}  \langle
	\Psi_\beta(\vec x
	,t) \bar\Psi_\alpha (0) \rangle\, ,\\
	C_3^q(t,t_s,\vec{q}) &=  \Gamma_{\alpha \beta} \sum\limits_{\vec x,\vec
	y} e^{i\vec q \vec y} \langle
	\Psi_\beta (\vec x,t_s) \mathcal{O}^q(\vec y,t) \bar\Psi_\alpha(0) \rangle
	\,
	\label{eq:cor},
\end{align}
where the nucleon interpolating operator is given by
\begin{align}
	\Psi_\alpha(x) = \epsilon_{abc}
                \left(\tilde{u}^T_a(x)C\gamma_5\tilde{d}_b(x)\right)\tilde{u}_{c,\alpha}(x)\,
		,
	\label{eq:ninterp}
\end{align}
and the operator of the local scalar current for a quark of flavor $q$ reads
\begin{align}
	O^q(x)&=\bar{q}(x)\mathbb{I} q(x).
\end{align}
The quark fields are Gaussian-smeared \cite{Gusken:1989ad} 
\begin{align}
	\tilde{q} = (1 + \kappa_{\rm G}\Delta)^{N_{\rm G}} q\,, \qquad q=u,d,
	\label{eq:gauss_smear}
\end{align}
using spatially APE-smeared  gauge links in the covariant Laplacian $\Delta$ \cite{APE:1987ehd}.

 \begin{figure}[t]
 \begin{center}
	 \includegraphics{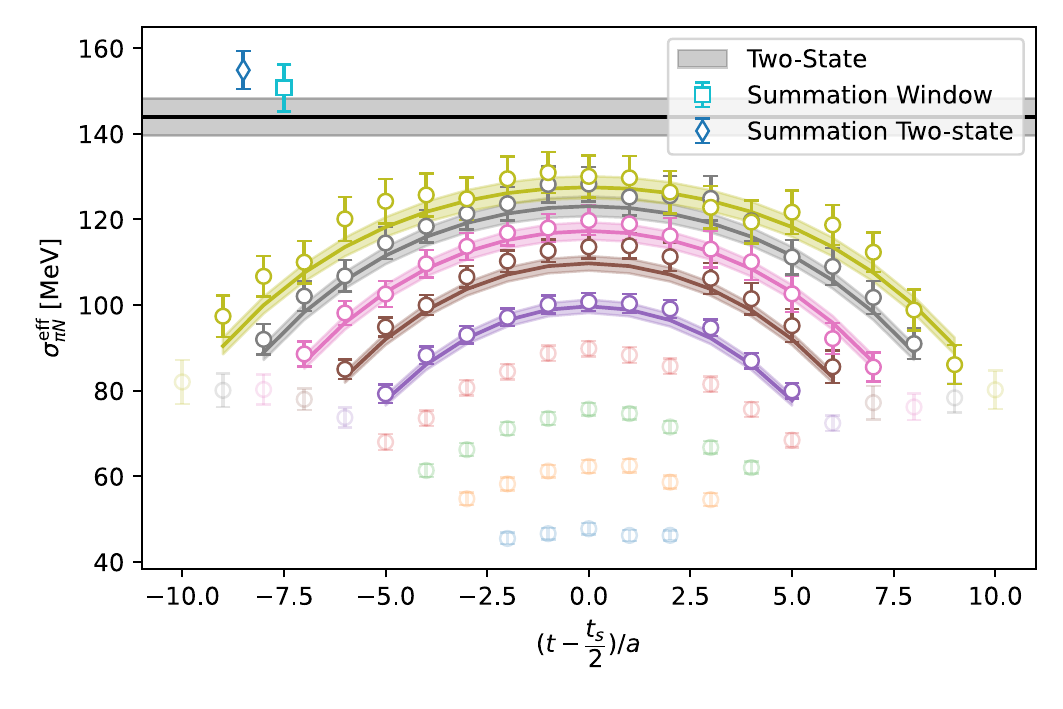}
	 \caption{Comparison of the extracted values for the summation window,
	 summation two-state, and explicit two-state analysis on ensemble N451.
 The data points show the effective form factor for all insertion times of the
 current between source and sink. The colored bands are the result of the two-state fit
 using Eq.~(\ref{eq:corrtwostate}), while the grey band corresponds to the
 extracted ground-state matrix element.}
 \label{fig:eff}
 \end{center}
 \end{figure}

For the disconnected part we use a variation of the
frequency splitting method \cite{Giusti:2019kff}, combining the
one-end-trick \cite{McNeile:2006bz}, generalized hopping parameter
expansion \cite{Gulpers:2013uca} and
hierarchical probing \cite{Stathopoulos:2013aci} (for more details see
Ref.~\cite{SanJose:2022yad}) to estimate the quark loops. The disconnected three-point function is given in terms of the quark loop and
the nucleon two-point function
\begin{align}
	C_3^{q,\rm disc} (t,t_s;\vec 0) = \langle L^{q}(\vec 0,
	t) \hat C_2(t_s,\vec p = \vec 0 )\rangle,
	\label{eq:disc_3pt}
\end{align}
where $L^q$ denotes the trace over quark loops of flavor $q$. For the
forward matrix element, the vacuum expectation of the loop and the two-point
functions need to be subtracted. We reduce the variance for two- and three-point
functions applying 
 the truncated solver method with bias correction
 \cite{Bali:2009hu,Blum:2012uh,Shintani:2014vja}. For details on the setup and
 the ensembles used see Ref.~\cite{Djukanovic:2022wru}.

 The scalar form factor may be extracted as the ratio of three- and two-point
 functions
\begin{align}
	{\rm Re}\,\frac{C_3^q(t,t_s)}{C_2(t,t_s)} &\xrightarrow{t,(t_s-t) \gg 0}
	G^q_{\rm S}\, ,
	\label{eq:ratio}
\end{align}
in the asymptotic limit of large time separations. The pion-nucleon sigma term
is a linear combination of the asymptotic ratios and light quark masses. It is
convenient to build an effective form factor, the asymptotic limit of which coincides with $\sigma_{\pi N}$. In
Fig.~\ref{fig:eff}, we show such an effective form factor for the ensemble N451.
In the region of large time separation to either source or sink one expects a
plateau, where the ground state dominates. The lack of a clear plateau region,
i.e. the curvature in Fig.~\ref{fig:eff},
already hints at sizeable effects due to excited states in this quantity.

 \begin{figure}[t]
	 \begin{center}
		 \includegraphics[width=.9\textwidth]{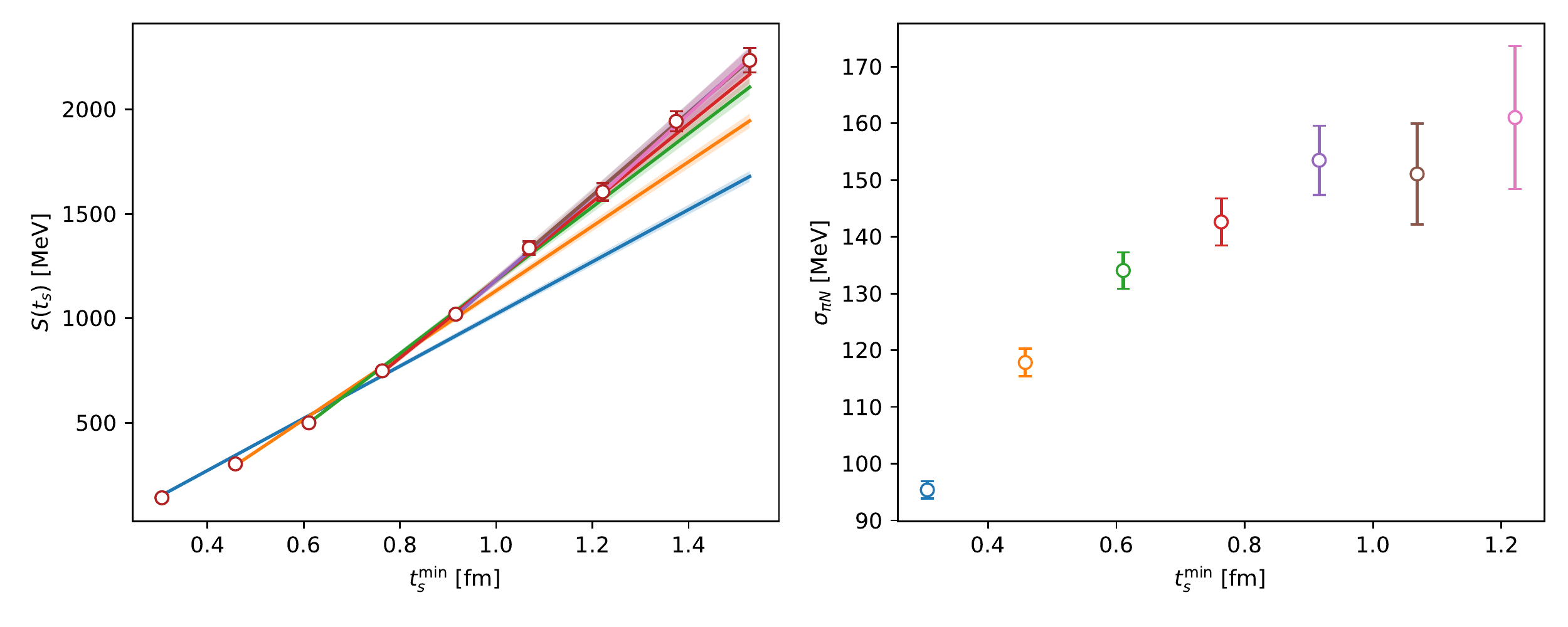}
		 \caption{Left: Linear fits to the summed
			 correlator (data points) on ensemble N451, i.e.
			 Eq.~(\ref{eq:sumcorra})
			 with $c_{10}=c_{11}=0$, for different starting
			 source-sink separations. The right panel shows  the
			 corresponding values for $\sigma_{\pi N}$ for the
			 different fits, where the color of the data points
		 match the color of the corresponding fit in the left plot.}
		 \label{fig:linfit}
	 \end{center}
	 
 \end{figure}
 \section{Excited States}\label{sec:esc}
 As a consequence of the unfavorable signal-to-noise ratio for matrix elements
 of baryonic
 operators \cite{Lepage:1989hd}, most calculations in the baryon sector suffer from contamination
 due to excited states (see \cite{Djukanovic:2021qxp} and references therein). We therefore have to include effects of excited states into the
 analysis of the correlators. In order to assess the remnant excited-state
 contribution, we compare the extraction based directly on the ratio of correlators and
 alternatively from the summed correlator. The latter is less sensitive to
 contributions from excited states, as these are parametrically more
 suppressed \cite{Maiani:1987by,Dong:1997xr,Capitani:2012gj,Green:2014xba}. We define the summed correlator of the effective form factor
 \begin{align}
	 S(t_s)=\sum\limits_{t=a}^{t_s-a} G_S^{\rm eff} (t,t_s).
	 \label{eq:sumcorr}
 \end{align}
 Including one excited state in the summed correlator, its asymptotic behavior
 reads
 \begin{align}
	 S(t_s) &=  (G_S + c_{11} e^{-\Delta t_s} ) (t_s-1) +2 c_{10} e^{-\Delta
	 t_s/2} {\rm csch} \frac{\Delta}{2} \sinh\frac{1}{2} (t_s-1) + \dots ,
	 \label{eq:sumcorra}
 \end{align}
 where $\Delta$ denotes the energy gap between ground and first excited state,
 and
 $c_{10}, \ c_{11}$ are the overlaps of the scalar current with ground-excited, and
 excited-excited states, respectively. Alternatively one may directly fit the effective
 form factor using an explicit two-state ansatz
 \begin{align}
	 G_S^{\rm eff} &=  G_S + c_{10}  e^{-\Delta t } +c_{10} e^{-\Delta
	 (t_s -t) }+ c_{11} e^{-\Delta t_s}.
	 \label{eq:corrtwostate}
 \end{align}

 \begin{figure}[t]
	 \begin{center}
		 \includegraphics[width=0.6\textwidth]{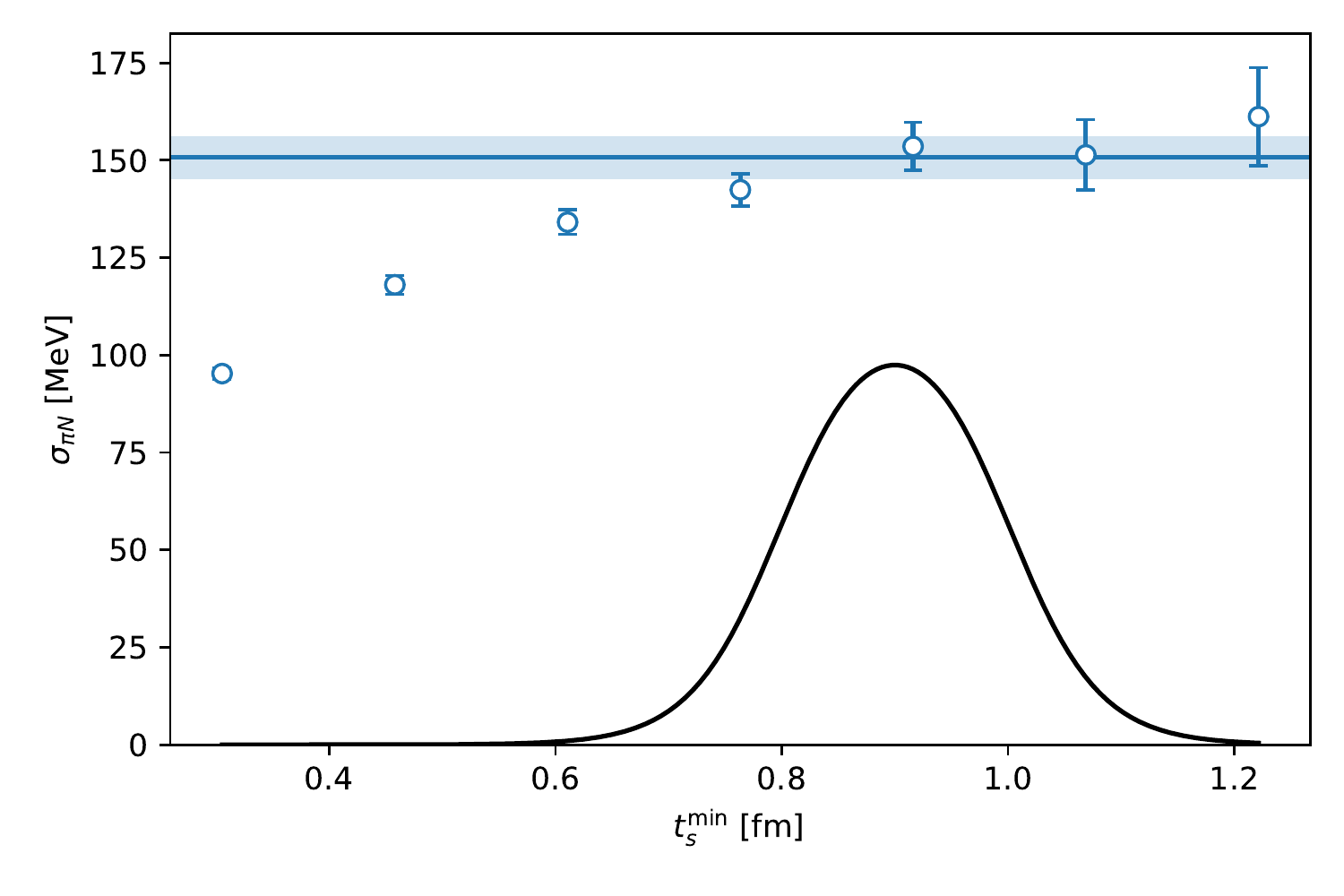}
		 \caption{Estimates for the pion-nucleon sigma term extracted via the summation
method, plotted as a function of the minimum source-sink separation
$t_s^{\rm min}$.
			 The blue shaded area corresponds to the window average
			 using the weights of Eq.~(\ref{eq:wts}) for the summation
			 extraction at different starting time slices $t_s^{\rm
			 min}$ (blue data
			 points) for ensemble N451. The black line shows the profile of the
		 weights for the parameters $t_s^{\rm min}$.}
	 \end{center}
 \end{figure}

We extend the
 available source-sink separations to include smaller values, down to $t_s=4$ in
 lattice units, see Ref.~\cite{Djukanovic:2022wru}. This enables us to
 monitor the stability of the  extraction of the matrix element based on a linear fit to the
 summed correlator, i.e. assuming ground state dominance. 

\begin{figure}[t]
	 \begin{center}
	 \includegraphics[width=.8\textwidth]{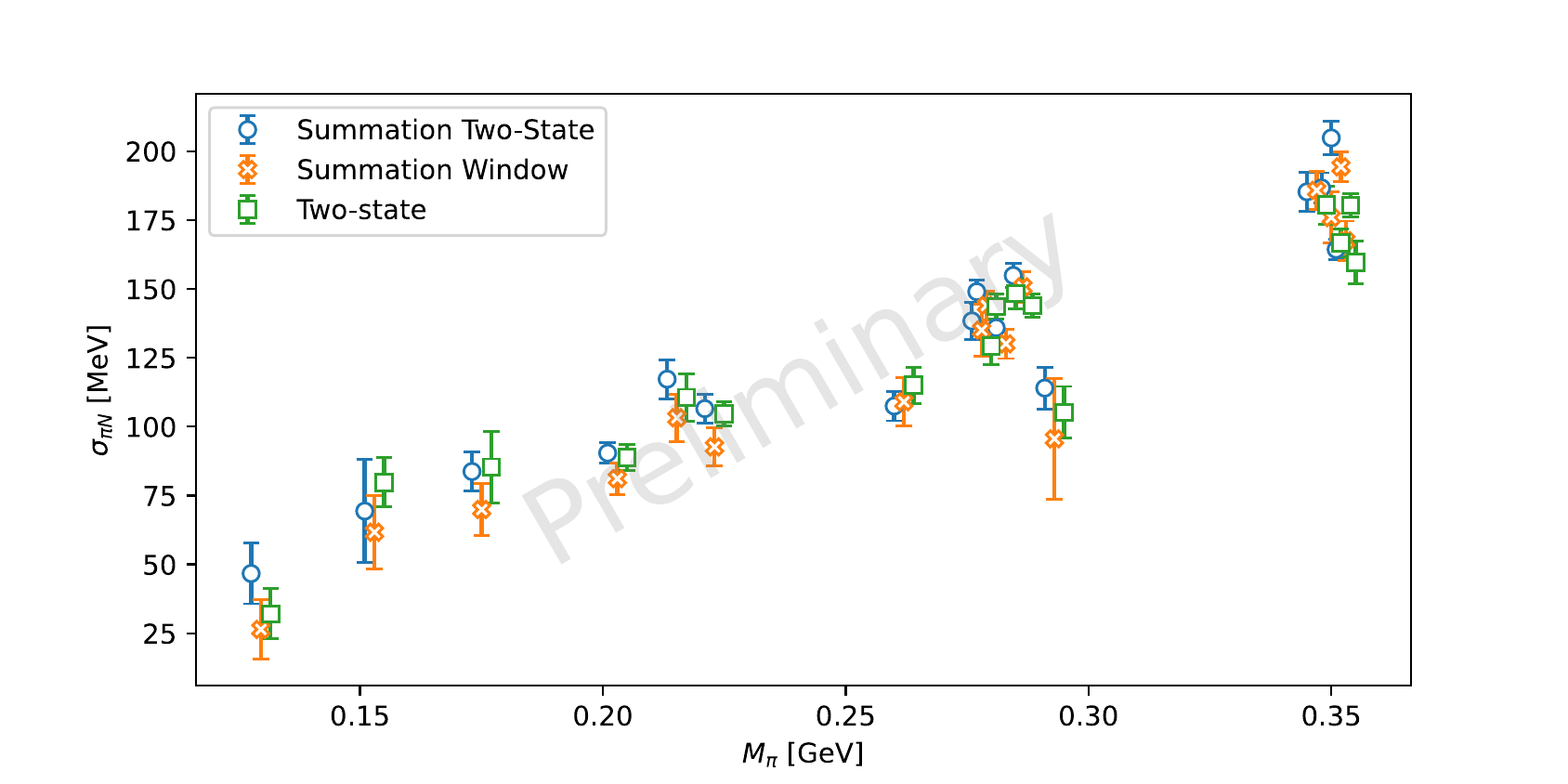}
	 \caption{Comparison of the extracted values for $\sigma_{\pi N}$ on all
		 ensembles using
		 the window average of the summed correlator (orange crosses),
		 the two-state ansatz for the summed correlator (blue
			 circle) and the direct two-state ansatz 
		 (green squares). }
	 \label{fig:compextract}
 \end{center}
 \end{figure}

 In Fig.~\ref{fig:linfit} we show the
 linear fit of Eq.~(\ref{eq:sumcorra}) as a function of the starting time
 slice $t_s^{\rm min}$ for ensemble N451. The corresponding values for the sigma-term are shown in the right
 panel, where the onset of a  plateau in the extraction is visible starting
 around 0.9 fm. We see that for very small $t_s^{\rm min}$ the extraction is still biased by the influence from excited states, while  for large $t_s^{\rm min}$ the error increases. With this in mind, instead of choosing one
 particular $t_s^{\rm min}$ to quote a final value for the summation method, we perform
 averages based on weights (see Ref.~\cite{Djukanovic:2022wru})
 \begin{align}
	 w_i &=  \frac{1}{2} \tanh \frac{t_s -t_{\rm lo} } {\delta t } -
	 \frac{1}{2} \tanh \frac{t_s -t_{\rm up}}{\delta t }.
	 \label{eq:wts}
 \end{align}
 These weights effectively define a window that suppresses unreliable estimates at very
 early starting times and, at the same time, reduces the influence of the noisy
 estimates at larger $t_s^{\rm min}$.  The parameters for the window are fixed
 in physical units and
 applied uniformly to all
 ensembles. We find 
 \begin{align}
	 t_{\rm lo} = 0.8\ {\rm fm}, \ \ t_{\rm up} = 1.0\ {\rm fm}, \ \ {\rm
	 and } \ \ \delta t=0.08 \ {\rm fm} 
	 \label{eq:wtpars}
 \end{align}
 leads to a stable set of estimates for the ground state matrix element.
 In addition, we fit the summed correlator using the ansatz of
 Eq.~(\ref{eq:sumcorra}), i.e. including excited states. However, the term
 corresponding to the excited-excited contribution $c_{11}$, as well as the
 energy gap between ground and first excited state $\Delta$ are not well
 constrained. We therefore choose a simplified ansatz, removing $c_{11}$ and, to
 stabilize the fits,
 apply Gaussian priors for the energy gap $\Delta$, with a central value
 corresponding to two non-interacting pions on the respective ensemble with a width of 5
 \%. The same caveats apply to the direct two-state fit ansatz of
 Eq.~(\ref{eq:corrtwostate}). In Fig.~\ref{fig:compextract} we compare the different extractions,
 and observe agreement within 2 standard deviations.

\section{Chiral and Continuum Extrapolation}
\begin{figure}[t]
	\begin{center}
	\includegraphics[width=0.45\textwidth]{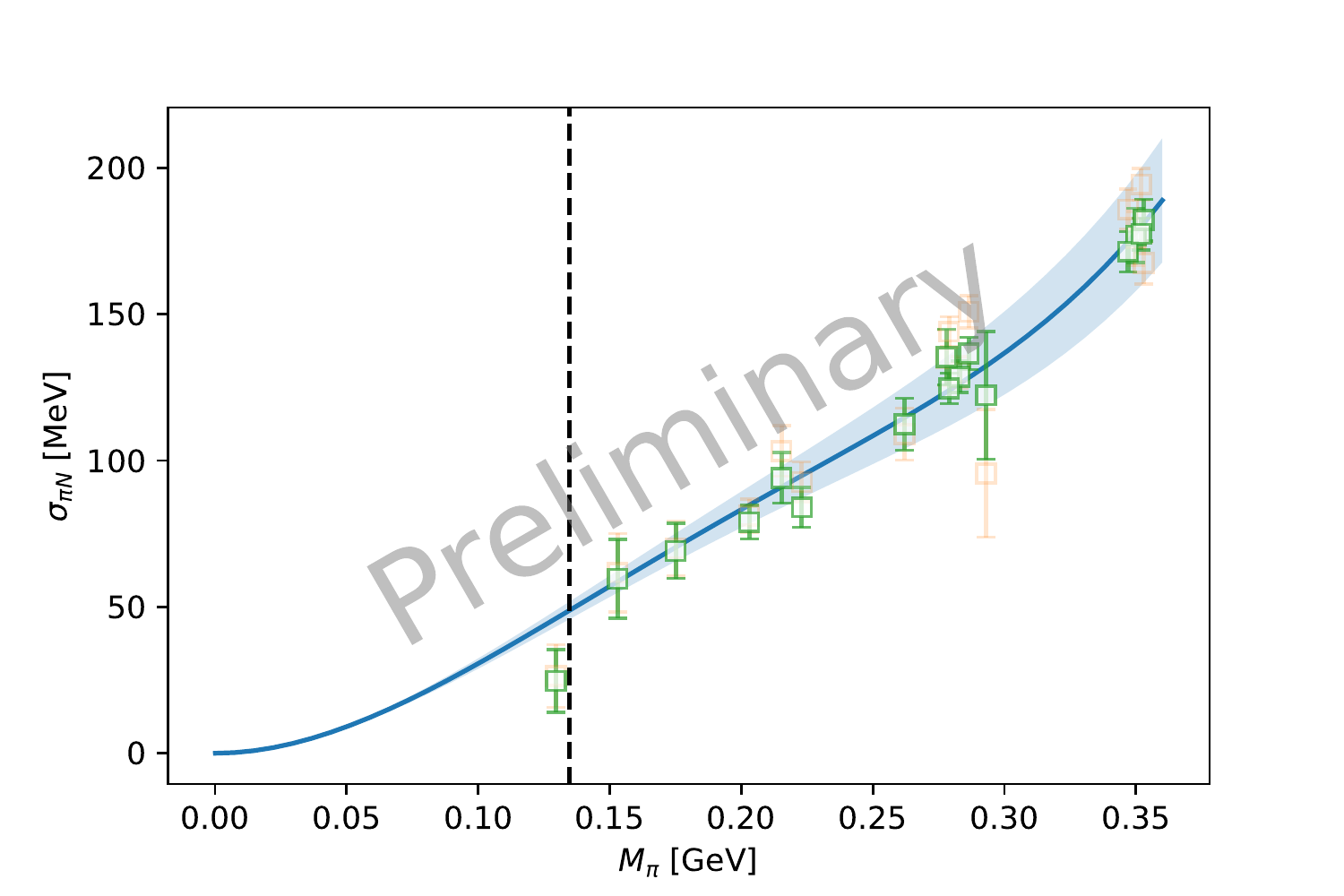}
	\includegraphics[width=0.41\textwidth]{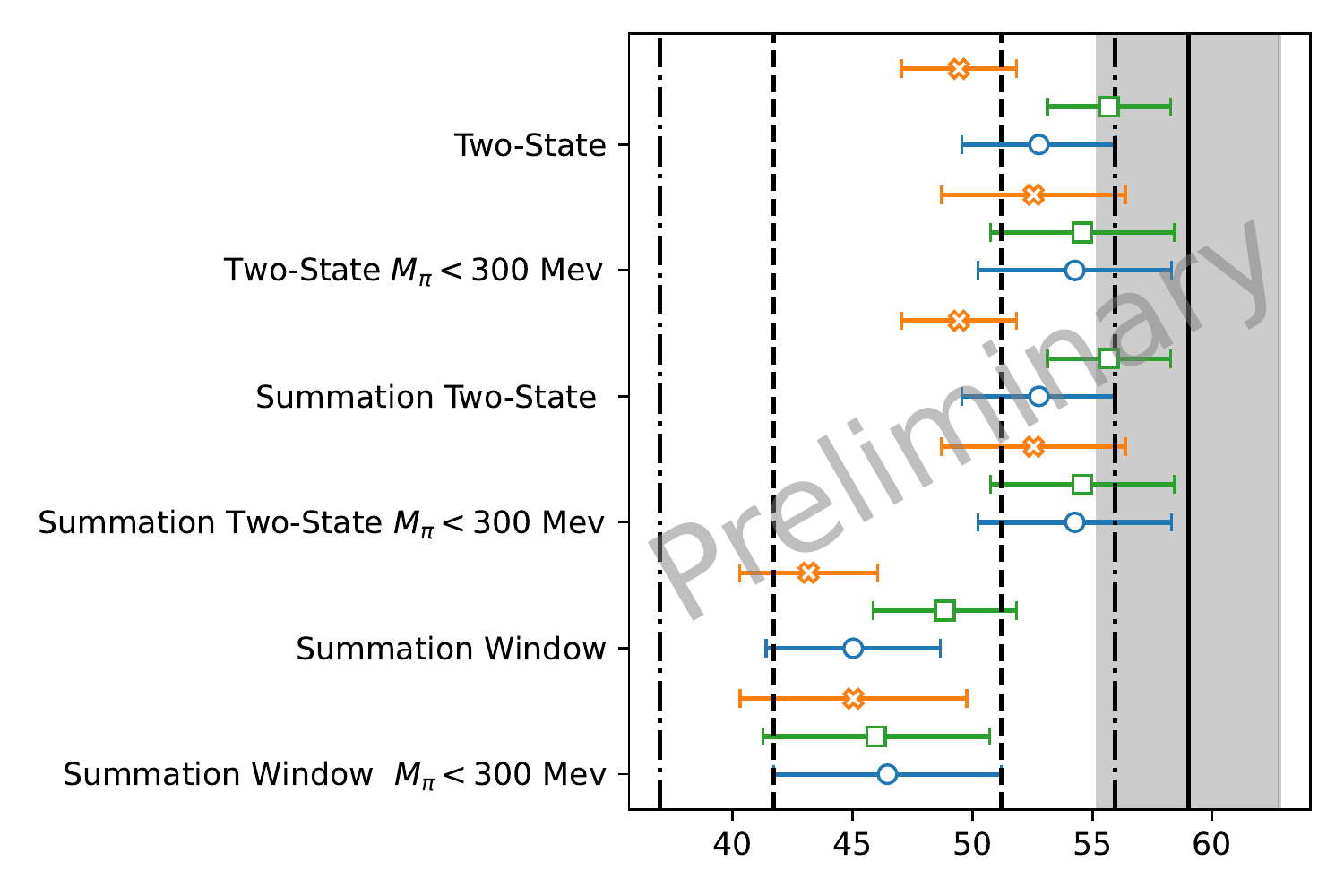}
	\caption{Left: Fit using the full form of
		Eq.~(\ref{eq:chpt}), without a cut in the pion mass on the window
	averaged summation data points in green/light orange, with/without corrections due to lattice spacing and
finite volume applied. Right: Compilation of results obtained  using the three
variations of the ChPT expressions, i.e. blue points, green square and orange
cross, correspond to  third order fits, including the full form, dropping the
chiral log, respectively, on all data sets, and with pion mass cuts as
indicated in the label. The grey band corresponds to the dispersive analysis
of
Ref. \cite{Hoferichter:2015dsa}.}
	\label{fig:ccf}
\end{center}
\end{figure}

The chiral expansion of the nucleon mass is known to sixth order in the chiral
counting \cite{Schindler:2006ha}. However, given the rapid growth in the number of coupling constants,
we restrict ourselves to the SU(2) expression at fourth chiral order,
which,
amended with terms parametrizing lattice spacing dependence and finite volume
corrections, reads 
\begin{align} \sigma_{\pi N} &=  (k_1+ k_a a) M_\pi^2 + k_2
M_\pi^3 +2 k_3 M^4 \log \frac{M_\pi}{\mu}+k_4 M_\pi^4 + k_L M_\pi^2
	\Bigl[\frac{1}{L} - \frac{M_\pi}{2}\Bigr] e^{-M_\pi L}.  \label{eq:chpt}
\end{align} 
The functional form of the finite volume corrections is taken from  \cite{Beane:2004tw}.
The coefficients $k_1,k_2$ and $k_3$ depend on known low-energy constants
(LECs), while $k_4$ receives contributions from a less well known fourth order
LEC and contributions from the mesonic Lagrangian at fourth order. Leaving all couplings free leads to unstable fits, especially after
applying cuts in the pion mass. We may stabilize the fits by either dropping
terms, noting that otherwise large cancellation happen between fourth order terms,
or fixing some of the coefficients to the values known from ChPT. Instead of
completely fixing the values we choose to apply Gaussian priors on the LECs,
i.e.  
\begin{align} k_1 &=  -4 c_1 = (4.44 \pm 0.12) \ {\rm GeV}^{-1} , \\ k_2 &=
	-\frac{9 g_A^2 }{64 \pi F_\pi^2} = (-8.52\pm 0.04) \ {\rm GeV}^{-2} ,\\
	k_3 &= -\frac{3}{32 \pi^2 F_\pi^2 } \Bigl( \frac{g_A^2}{m_n} - 8 c_1 +
	c_2 +4 c_3 \Bigr) - \frac{c_1}{8 \pi^2 F_\pi^2} = (-11.38 \pm 0.35) \
{\rm GeV}^{-3}, \label{eq:priorschpt} 
\end{align} where the values for the LECs
are taken from \cite{Hoferichter:2015hva}.  The left panel of Fig.~\ref{fig:ccf} 
shows one particular fit, using the full form of Eq.~(\ref{eq:chpt}) on summation
window averaged data on all ensembles. We further analyze variations of Eq.~(\ref{eq:chpt}) in
order to assess the relevance of individual terms to the final result.
In the right panel of Fig.~\ref{fig:ccf} we compare results from three types of
ChPT ans\"atze: (1) using ChPT including terms up to third order, (2)
dropping the chiral log, i.e. k3 = 0 and (3) the full expression of
Eq.~(\ref{eq:chpt}). Each ansatz is applied to the different data sets as
explained in Sec.~\ref{sec:esc}, either with a cut of 300 MeV in the pion mass, or
over the full pion mass range. In general the variations for the
fit functions agree quite well within each set, while for the data sets with an explicit
two-state ansatz including priors for the energy gap, a systematic shift towards larger
values of $\sigma_{\pi N}$ is visible. However, the errors are still quite large
and all extractions agree within two standard deviations. Nevertheless it is evident that excited
states play an important role for the scalar matrix element.

We intend to further increase statistics,
especially for the disconnected part, and perform fits based on SU(3) ChPT
expressions, in order to have a handle on the strange sigma term  $\sigma_s$.
Furthermore, including constraints from the nucleon mass dependence might prove helpful to further
constrain the sigma terms.

\section{Acknowledgments} 
This work was supported in part by the European
Research Council (ERC) under the European Union’s Horizon 2020 research and
innovation program through Grant Agreement No.\ 771971-SIMDAMA and by the
Deutsche Forschungsgemeinschaft (DFG) through the Collaborative Research Center
SFB~1044 ``The low-energy frontier of the Standard Model'', under grant
HI~2048/1-2 (Project No.\ 399400745) and in the Cluster of Excellence “Precision
Physics, Fundamental Interactions and Structure of Matter” (PRISMA+ EXC 2118/1)
funded by the DFG within the German Excellence strategy (Project ID 39083149).
Calculations for this project were partly performed on the HPC clusters
``Clover'' and ``HIMster2'' at the Helmholtz Institute Mainz, and ``Mogon 2'' at
Johannes Gutenberg-Universit\"at Mainz.  The authors gratefully acknowledge the
Gauss Centre for Supercomputing e.V. (www.gauss-centre.eu) for funding this
project by providing computing time on the GCS Supercomputer systems JUQUEEN and
JUWELS at J\"ulich Supercomputing Centre (JSC) via grants CHMZ21, HMZ23,NUCSTRUCLFL, GCSNUCL2PT and CHMZ36
(the latter through the John von Neumann Institute for Computing (NIC)), as well
as on the GCS Supercomputer HAZELHEN at H\"ochstleistungsrechenzentrum Stuttgart
(www.hlrs.de) under project GCS-HQCD.
  
Our programs use the QDP++ library~\cite{Edwards:2004sx} and deflated SAP+GCR
solver from the openQCD package~\cite{Luscher:2012av}, while the contractions
have been explicitly checked using~\cite{Djukanovic:2016spv}. We are grateful to
our colleagues in the CLS initiative for sharing the gauge field configurations
on which this work is based.

\bibliographystyle{JHEP}
\bibliography{bib}

\end{document}